\documentclass[12pt,a4paper]{article}
\usepackage[a4paper,hmargin={1.5cm,1.5cm},vmargin={2cm,2cm}]{geometry}
\usepackage[utf8]{inputenc}
\usepackage[T1]{fontenc}
\usepackage[affil-sl]{authblk}

\usepackage[dvipsnames, table]{xcolor}
\usepackage[singlelinecheck=false]{caption}

\usepackage[linesnumbered,ruled]{algorithm2e}
\usepackage[colorlinks=true,linkcolor=Rhodamine,citecolor=Rhodamine]{hyperref}

\usepackage{adjustbox}
\usepackage[thinlines]{easytable}
\usepackage{threeparttable}

\usepackage{array}
\usepackage{rotating}
\usepackage{enumitem}

\usepackage{titlesec}
\usepackage[square, numbers, sort&compress]{natbib}
\usepackage{url}
\usepackage{hyphenat}
\usepackage{parskip}

\usepackage{multirow}
\usepackage{tabularray}
\usepackage{tabularx}
\usepackage{makecell}
\usepackage{amsmath}
\usepackage{upgreek}
\usepackage{breqn}
\usepackage{mathtools}
\usepackage{amsfonts}
\usepackage{amscd}
\usepackage{amssymb}

\usepackage{tikz}
\usepackage{graphicx}
\usepackage{pifont}
\usepackage{mathrsfs}
\usepackage{fancyhdr}
\usepackage{subfiles}

\usepackage{amsthm}

\definecolor{primary}{HTML}{6200ea}
\definecolor{secondary}{HTML}{00e5ff}

\begin{document}
\font\titleFont=cmr12 at 17pt
\title{{\titleFont An Empirical Investigation of Reconstruction-Based Models for Seizure Prediction from ECG signals}}

\def\correspondingauthor{\footnote{~Corresponding Author: fghaderi@modares.ac.ir (F. Ghaderi)}}

\author{Mohammad Reza Chopannavaz}
\author{Foad Ghaderi \correspondingauthor{}}

\affil{Human-Computer Interaction Lab., Faculty of Electrical and Computer Engineering, Tarbiat Modares University, Tehran, Iran}

\date{} 

\maketitle

\section*{Abstract}\label{sec_absract}
Epileptic seizures are transient neurological events characterized by abnormal and excessive neuron activity in the brain, which are often associated with measurable disturbances in the cardiovascular system. Traditionally, electroencephalogram (EEG) signals have served as the primary modality for seizure prediction due to their direct measurement of brain activity and high diagnostic precision. However, their cost, sensitivity to noise, and practical deployment constraints limit their applicability outside controlled clinical environments. To overcome these challenges, recent studies have increasingly investigated electrocardiogram (ECG) signals as a practical and non-invasive alternative for seizure prediction in real-world settings. Evidence suggests that ECG-derived cardiac signatures may precede clinical seizure onset, offering a viable window for early detection. In this paper, we propose a reconstruction-based anomaly detection framework that integrates time-frequency representations with advanced deep learning models to capture deviations in heart rate dynamics associated with seizure onset. Afterward, reconstruction error is smoothed, and an adaptive thresholding strategy is applied to reduce false alarms. The method was evaluated on the Siena database, achieving a specificity of 99.16\%, accuracy of 76.05\%, and a false positive rate (FPR) of 0.01/h, with an average prediction horizon of 45 minutes prior to seizure onset. These results demonstrate that ECG-based prediction can provide clinically actionable early warnings while improving patient accessibility and comfort. Nevertheless, this performance reflects a trade-off favoring high specificity over sensitivity, resulting in reduced FPR and aligning with clinical requirements for reliable deployment.

\textbf{Keywords}~
Anomaly Detection; Autoencoders; Deep Learning; Electrocardiogram; Epilepsy; Seizure Prediction.

\section{Introduction}\label{sec_intro}
Epilepsy is a chronic neurological disorder defined by recurrent, unprovoked seizures arising from abnormal electrical activity in the brain. These events vary considerably in their occurrence, frequency, and severity, posing significant challenges for reliable diagnosis and prediction. Despite advances in therapeutic interventions, a substantial proportion of patients continue to experience recurrent episodes following treatment--a condition referred to as drug-resistant or intractable epilepsy \cite{Dalic.2016}. Consequently, many individuals fail to achieve adequate seizure control, leading to diminished quality of life, increased risk of injury, restrictions in daily activities, and a higher prevalence of psychological comorbidities \cite{Fisher.2014, DeBoer.2008}.

Given the inherently unpredictable nature of epilepsy, the development of reliable seizure prediction and management strategies remains a critical research priority. Seizures typically evolve through distinct phases, including the inter-ictal (between seizures), pre-ictal (preceding seizure onset), ictal (during the seizure), and post-ictal (recovery) stages \cite{Ufongene.2020}. Among these, the inter-ictal phase reflects baseline physiological conditions and can therefore serve as a reference for identifying deviations associated with the pre-ictal state. The ability to accurately anticipate an impending seizure would enable timely interventions, thereby improving patient safety, independence, and overall quality of life.

Traditionally, seizure prediction approaches have primarily relied on electroencephalogram (EEG) recordings to identify abnormal neural activities \cite{Jaishankar.2023,MohammadkhaniGhiasvand.2020}. However, practical limitations such as limited portability, discomfort during long-term use, and vulnerability to motion artifacts have motivated the exploration of alternative biomarkers. One such alternative is electrocardiogram (ECG) signals, which have emerged as a promising non-invasive and more accessible option. Recent studies indicate that ECG-based approaches can capture meaningful physiological changes associated with the pre-ictal phase \cite{Manni.2020}. Furthermore, recent reviews highlight the growing success of machine learning techniques applied to ECG data, positioning them as a scalable and effective alternative to traditional EEG-based methods \cite{Chopannavaz.2026, Mason.2024}.

The growing interest in ECG-based seizure prediction is largely grounded in physiological evidence demonstrating that epileptic seizures disrupt the autonomic nervous system (ANS), leading to observable alterations in cardiac activity. During seizures, autonomic imbalance can induce abrupt fluctuations in heart rate, manifesting as tachycardia, bradycardia, arrhythmias, and in severe cases, transient asystole. Importantly, these cardiac disturbances are not limited to the ictal phase, as they may also emerge during the pre-ictal period, where subtle variations in heart rate and other ECG-derived features have been identified as potential early indicators of impending seizures \cite{Poh.2012, Zijlmans.2002}. Such pre-ictal cardiac signatures provide a promising, non-invasive pathway for seizure prediction, particularly in scenarios where conventional EEG monitoring is impractical due to its limited portability and operational constraints.

Despite this strong physiological basis, the detection of pre-ictal cardiac alterations remains a challenging task. These changes are often subtle, non-stationary, and distributed across multiple temporal scales. Autonomic influences may appear as gradual shifts in heart rate variability (HRV), transient waveform irregularities, or localized spectral changes, patterns that are not adequately captured by single-domain or fixed-resolution signal representations. Consequently, effective characterization of pre-ictal dynamics necessitates analytical approaches capable of jointly capturing temporal and spectral information. In parallel, practical seizure prediction systems prioritize high specificity to minimize false alarms, even at the expense of reduced sensitivity, due to the critical impact of alarm fatigue on real-world clinical usability. This constraint underscores the need for robust and reliable detection strategies. To address the complexity of pre-ictal dynamics, time-frequency analysis provides a natural framework, enabling multi-scale representation of ECG signals and improving the detection of subtle physiological variations associated with autonomic dysregulation relative to baseline cardiac activity.

Numerous studies have demonstrated the potential of leveraging heart rate dynamics for epileptic seizure prediction. In particular, these studies have focused on HRV as a key indicator of ANS dysregulation. Prior works \cite{Hashimoto.2013, Hashimoto.2013a, Fujiwara.2014, Behbahani.2016, Gagliano.2020, Leal.2021} have shown that HRV features across multiple domains--including time, frequency, and nonlinear measures--can reveal patterns associated with impending seizure onset. These alterations reflect underlying shifts in sympathetic and parasympathetic activity, which are commonly observed during the pre-ictal phase. Collectively, these findings highlight the promise of HRV-based analysis as a non-invasive approach for seizure prediction. Nevertheless, HRV-based methods have drawbacks that affect their applicability in real-time settings. A primary challenge is the requirement of sustained data windows, typically two to three minutes in duration, to calculate reliable HRV metrics. This dependence introduces inherent latency in feature extraction, making such methods less suitable for immediate seizure prediction and timely intervention in dynamic, real-world environments.

Taking into account these timing constraints, Ode et al. \cite{Ode.2022}, developed a rapid-response model using a Self-Attentive Autoencoder (SA-AE) designed for detecting anomalies in heart rate patterns with minimal latency. By focusing on reconstruction error as an anomaly score, their SA-AE model enables quick identification of pre-ictal patterns without the extensive preprocessing required for conventional HRV feature extraction. Their results demonstrated that deep learning architectures like the SA-AE are highly effective at detecting early signs of seizures from physiological data, thereby enabling more responsive seizure prediction.

Building upon this line of work, this study extends ECG-based anomaly detection by modeling cardiac dynamics across multiple temporal and spectral scales. While reconstruction-based autoencoders are well-suited for identifying deviations from normal physiological behavior, their effectiveness is closely tied to the representation of the input signal. Traditional HRV features rely on statistical summaries computed over extended time windows, which may obscure transient or evolving physiological changes and introduce delays. In contrast, time-frequency representations preserve both temporal localization and spectral content, enabling continuous and fine-grained characterization of cardiac activity without the need for prolonged aggregation. By capturing instantaneous variations directly from the ECG waveform, such representations provide a richer physiological context for learning patient-specific baseline dynamics. Within this framework, seizure prediction is formulated as an anomaly detection problem, meaning models are trained on inter-ictal data to learn normal cardiac behavior, and impending seizures are anticipated when reconstruction errors increase due to deviations associated with pre-ictal autonomic dysregulation.

In response to the critical challenges of seizure prediction, this study introduces the following significant contributions:
\begin{itemize}
	\item We formulate ECG-based seizure prediction as a reconstruction-based anomaly detection problem, enabling fully unsupervised modeling of pre-ictal cardiac abnormalities without reliance on specific annotations or handcrafted feature engineering.

	\item We propose a multi-scale time-frequency reconstruction-based framework for ECG analysis, and systematically investigate the effectiveness of different deep learning architectures in capturing temporal and spectral dependencies of pre-ictal cardiac dynamics.
\end{itemize}

The remainder of this paper is organized as follows. Section \ref{sec_materials_methods} presents the key components and methods used to construct the proposed framework. Section \ref{sec_postproces} describes the post-processing procedures, including reconstruction error smoothing and adaptive thresholding. Section \ref{sec_evaluation} introduces the evaluation methodology and performance metrics. Section \ref{sec_results} presents the experimental results and analyzes the impact of key design choices, including segmentation strategies, feature representations, and model architectures. Section \ref{sec_previous_works} compares the proposed framework with existing ECG-based seizure prediction methods. Finally, Sections \ref{sec_discussion} and \ref{sec_conclusion} present the discussion and conclusion, respectively.

\section{Materials and Methods} \label{sec_materials_methods}
\begin{figure}[!ht]
	\begin{center}
		\includegraphics[width=\textwidth-3cm]{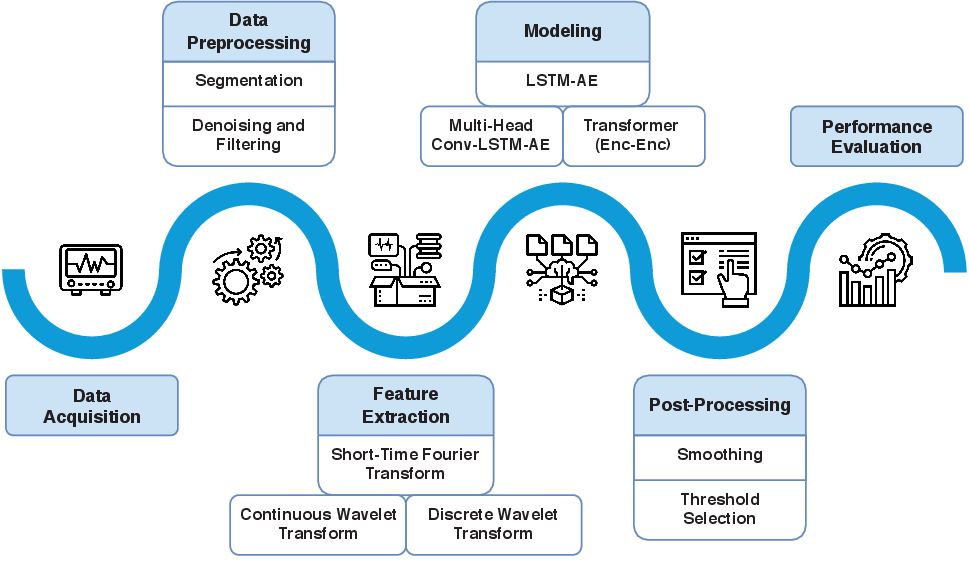}
		\caption{Overall components of the proposed framework for Epileptic Seizure Prediction.}
		\label{fig_process}
	\end{center}
\end{figure}

The proposed seizure prediction framework, illustrated in Fig. \ref{fig_process}, follows a structured pipeline including data acquisition, preprocessing, feature extraction, modeling, post-processing, and performance evaluation. Initially, raw ECG signals are collected and preprocessed to ensure signal quality and consistency. Subsequently, time-frequency transformation techniques are applied to obtain multi-resolution representations of ECG signals. These representations are then processed using multiple deep learning architectures within a reconstruction-based anomaly detection framework. Post-processing is subsequently applied to refine predictions by reducing noise and improving robustness. Finally, the performance of the proposed framework is evaluated using standard metrics.


\subsection{Data Acquisition} \label{sub_acquisition}
In this study, data were obtained from the Siena Scalp EEG Database \cite{Detti.2020}, a comprehensive open-access dataset provided by the University of Siena, Italy. The dataset includes simultaneous scalp EEG and ECG recordings from 14 patients with epilepsy, primarily diagnosed with focal epilepsy, with some cases involving generalized tonic-clonic seizures. Each recording contains at least one complete seizure event. A summary of patient demographics and selected clinical characteristics is presented in Table \ref{tbl_dataset}.

The ECG signals in the dataset are sampled at 512 Hz, providing sufficient temporal resolution to capture subtle variations in cardiac activity associated with impending seizures. In addition, the dataset includes expert annotations specifying seizure onset and offset times, enabling precise alignment of cardiac dynamics with seizures \cite{Detti.2020a, Goldberger.2000}.

\begin{table}[!ht]
	\centering
	\caption{Characteristics of the patients included in the study. In this table, IAS is focal onset impaired awareness; WIAS is focal onset without impaired awareness; FBTC is focal to bilateral tonic-clonic; T is temporal; F is frontal; R is right; L is left; and B is Bilateral.}
	\label{tbl_dataset}
	\renewcommand{\arraystretch}{1.15}
	\scalebox{0.89}{
		\begin{tabular}{cccccccc}
			Patient ID & Age & Gender & Seizure Type & Seizures \# & Recording (Min) \\ \hline
			PN00       & 55  & Male   & IAS          & 5           & 198             \\ \hline
			PN01       & 46  & Male   & IAS          & 2           & 809             \\ \hline
			PN03       & 54  & Male   & IAS          & 2           & 752             \\ \hline
			PN05       & 51  & Female & IAS          & 3           & 359             \\ \hline
			PN06       & 36  & Male   & IAS          & 5           & 722             \\ \hline
			PN07       & 20  & Female & IAS          & 1           & 523             \\ \hline
			PN09       & 27  & Female & IAS          & 3           & 410             \\ \hline
			PN10       & 25  & Male   & FBTC         & 10          & 1002            \\ \hline
			PN11       & 58  & Female & IAS          & 1           & 145             \\ \hline
			PN12       & 71  & Male   & IAS          & 4           & 246             \\ \hline
			PN13       & 34  & Female & IAS          & 3           & 519             \\ \hline
			PN14       & 49  & Male   & WIAS         & 4           & 1408            \\ \hline
			PN16       & 41  & Female & IAS          & 2           & 303             \\ \hline
			PN17       & 42  & Male   & IAS          & 2           & 308             \\ \hline
		\end{tabular}
	}
\end{table}

\subsection{Data Preprocessing} \label{sub_preprocess}
The preprocessing stage aims to enhance the quality of ECG signals through segmentation and denoising, preparing the data for feature extraction and model training. Initially, the continuous ECG signals are divided into manageable time windows. Subsequently, denoising techniques are applied to reduce noise and artifacts inherent in ECG recordings. These steps improve signal clarity and reliability, providing cleaner input for downstream analysis.

\subsubsection{Segmentation}
During segmentation, the ECG signal is partitioned into fixed-length time windows to facilitate the analysis of temporal patterns preceding seizure events. This step is essential for isolating meaningful segments within the continuous signal, enabling the identification of pre-ictal and ictal characteristics associated with seizure onset.

As a result, to capture the dynamic nature of cardiac activity, multiple window lengths--1, 5, and 10 seconds--are employed, along with different overlap configurations, including no overlap and partial overlaps of 1, 3, and 5 seconds. These configurations allow for systematic evaluation of the impact of window size and overlap on prediction performance and feature stability. Larger windows provide a broader temporal context, potentially capturing more stable heart rate trends, whereas shorter, overlapping windows, offer higher temporal resolution, which is critical for detecting rapid and transient pre-ictal changes. Moreover, overlapping segments preserve temporal continuity and reduce the risk of information loss at segment boundaries.

\subsubsection{Denoising and Filtering}
Denoising is a critical step in ECG signal processing, particularly for reliable feature extraction and seizure prediction. Common artifacts in ECG recordings include power line interference, motion artifacts, and baseline wander. Notably, baseline wander--often caused by respiration or body movement--may contain physiologically relevant information reflecting interactions between the ANS and cardiovascular dynamics during seizure development. Therefore, completely removing this component may lead to the loss of informative low-frequency patterns.

To address this, a low-pass filtering strategy is adopted to preserve slow-varying components, including baseline wander, while attenuating high-frequency noise that may obscure relevant signal features. This approach maintains a balance between noise reduction and information preservation, ultimately supporting more robust and accurate seizure prediction.

\subsection{Feature Extraction}
In order to accurately predict seizure onsets, reliable feature extraction is necessary to transform input ECG signals into meaningful representations that capture significant patterns, temporal trends, and characteristics relevant to seizure prediction. The intricate and high-dimensional nature of ECG signal necessitates extracting features that capture subtle pre-ictal dynamics, as these are vital for accurate and effective seizure prediction. To address this complexity, several techniques have been investigated. These include the set of features created by concatenating the coefficients derived from the Discrete Wavelet Transform (DWT), the scalogram representation generated by the Continuous Wavelet Transform (CWT), and the spectrogram representation produced by the Short-Time Fourier Transform (STFT). To assess whether each method was effective in improving signal representation and its impact on model performance, each method was investigated independently. In the following, we provide a detailed overview of each technique, the motivation behind the chosen parameters, and their significance in the context of seizure prediction.

\subsubsection{Discrete Wavelet Transform}
In the first approach, DWT was used to decompose the ECG signal into time-frequency components, enabling simultaneous analysis of transient and steady-state features. In contrast to Fourier Transform, DWT provides both time and frequency information, making it effective for ECG signals with abrupt changes. The DWT of a signal $x[n]$, sampled at discrete points, is computed using a mother wavelet $\psi(n)$, scaled and translated as follows:
\begin{equation}
	W_{j,k} = \sum_{n} x[n] \cdot \psi_{j,k}[n]
\end{equation}
where:
\begin{equation}
	\psi_{j,k}[n] = 2^{-j/2} \psi\left[2^{-j} n - k\right]
\end{equation}
Here, $j$ denotes the scale index, $k$ is the translation index, and $\psi(n)$ is the mother wavelet. These coefficients $W_{j,k}$ encapsulate both the temporal and spectral characteristics of the signal.

In this study, the sym4 wavelet was selected due to its symmetry and structural similarity to ECG waveforms. This will help in minimizing distortion while effectively capturing both rapid changes (e.g., arrhythmias) and smooth trends (e.g., HRV) \cite{Patro.2016, Phuphanin.2024}. The decomposition was performed up to level 3, balancing the preservation of high-frequency components, such as noise and rapid fluctuations, with low-frequency trends that reflect broader physiological changes \cite{Hatamian.2023}. The decomposition process at each level involved applying low-pass and high-pass filters, yielding approximation ($c_{A_j}$) and detail ($c_{D_j}$) coefficients:
\begin{equation}
	\begin{split}
		c_{A_j}[n] & = \sum_{k} h[k] \cdot c_{A_{j-1}}[2n-k] \\
		c_{D_j}[n] & = \sum_{k} g[k] \cdot c_{A_{j-1}}[2n-k]
	\end{split}
\end{equation}
where $h[k]$ and $g[k]$ denote the low-pass and high-pass filter coefficients, respectively. After decomposition, the coefficients $c_{A_3}, c_{D_3}, c_{D_2}, c_{D_1}$ were normalized to mitigate differences in signal magnitude across patients, ensuring comparability across datasets. Finally, the normalized coefficients were concatenated into a feature vector, creating a compact and robust representation of ECG signals for subsequent analysis.

\subsubsection{Continuous Wavelet Transform}
In the second approach, the CWT was employed to analyze the energy distribution of the ECG signal across both the time and frequency domains. Unlike traditional methods, CWT provides a detailed time-frequency representation, making it highly effective for detecting transient features in non-stationary biomedical signals such as ECG. The CWT of a signal $x(t)$ is defined as:
\begin{equation}
	C(a, b) = \int_{-\infty}^{\infty} x(t) \psi^*_{a,b}(t) \, dt
\end{equation}
where:
\begin{equation}
	\psi^*_{a,b}(t) = \frac{1}{\sqrt{a}} \psi\left(\frac{t-b}{a}\right)
\end{equation}
Here, $a > 0$ indicates the scale parameter, controlling frequency variations, while $b$ denotes the translation parameter, governing time localization. The function $\psi^*(t)$ is the complex conjugate of the mother wavelet. This decomposition enables precise analysis of signal variations at different time instances and frequency bands.

In this study, the Mexican hat wavelet (mexh) was selected as the mother wavelet due to its strong resemblance to QRS waves and excellent localization properties in both the time and frequency domains. The mathematical definition of the Mexican hat wavelet is given by:
\begin{equation}
	\psi(t) = \left(1 - t^2\right)e^{-t^2/2}
\end{equation}
This wavelet is particularly well-suited for identifying transient energy shifts in ECG signals, which are crucial for detecting pre-ictal activity preceding seizure events. Prior research has demonstrated that mexh effectively captures subtle energy variations in biomedical signals, making it a reliable tool for analyzing rapid transitions in ECG signals \cite{Burke.2004}.

To achieve an optimal balance between resolution and computational efficiency, we selected a scale range of $a = 128$. This range enables capturing both low-frequency trends (e.g., heart rate variations) and high-frequency changes (e.g., noise or sharp bursts). The scalogram, which represents the energy distribution of the signal across different scales and time instances, is computed as:
\begin{equation}
	E(a, b) = |C(a, b)|^2
\end{equation}
where $E(a, b)$ denotes the energy at scale $a$ and translation $b$.
Following the decomposition, to ensure consistency across datasets, scalogram energy was normalized, enhancing seizure prediction by identifying subtle pre-ictal patterns.

\subsubsection{Short-Time Fourier Transform}
In the third approach, the STFT was employed to analyze the time-frequency characteristics of ECG signals, capturing both steady-state trends and transient variations in ECG signals. These variations often include abrupt spectral energy shifts, which are critical indicators of pre-ictal states. The STFT of a signal $x(t)$ is defined as:
\begin{equation}
	X(t, f) = \int_{-\infty}^{\infty} x(\tau) w(\tau - t) e^{-j 2 \pi f \tau} d\tau
\end{equation}
where $w(\tau - t)$ denotes a window function that selects a segment of the signal centered at time $t$, and $f$ indicates the analyzed frequency component. This equation demonstrates that the signal is divided into short segments via the window function $w$, and the Fourier Transform is then applied to each segment, providing a localized frequency analysis.

In this study, a window size of 512 samples was selected to balance temporal and spectral resolution, ensuring the detection of both rapid transitions (e.g., arrhythmic events) and long-term variations (e.g., heart rate trends). A smaller window improves temporal resolution at the cost of frequency accuracy, while a larger window enhances frequency resolution but reduces time localization. This window size aligns well with the typical ECG sampling rate, effectively capturing both low and high-frequency features \cite{Huang.2019}.

Once the STFT coefficients are obtained, the spectrogram is computed by computing the squared magnitude of the STFT:
\begin{equation}
	S(t, f) = |X(t, f)|^2
\end{equation}
where $S(t, f)$ denotes the signal's energy distribution across time and frequency. This spectrogram provides a visual representation of how the power of different frequency components varies over time. Next, to ensure consistency across various patient datasets, the spectrogram energy was normalized, reducing the impact of inter-ictal variability while preserving the most relevant pre-ictal characteristics.

\subsection{Modeling} \label{sub_modeling}
In seizure prediction, the ability to detect anomalies in physiological signals is an essential component of early warning systems. These anomalies often manifest as subtle deviations from normal patterns during the pre-ictal phase, which precedes seizure onset. In this study, we present a reconstruction-based anomaly detection approach that can be used to predict the onset of epileptic seizures in an unsupervised manner. By analyzing signal representations produced through feature extraction, the approach minimizes reliance on labeled data, thereby enhancing its adaptability and scalability across diverse patient profiles.

The core idea of this study is based on the reconstruction error generated by models trained exclusively on normal segments of ECG signals. These models learn to represent the baseline patterns of the signal, capturing its inherent characteristics under typical, non-seizure conditions. However, during the pre-ictal phase, subtle yet significant deviations emerge in the signal's characteristics as the heart responds to physiological changes preceding a seizure event. These deviations, often undetectable through direct observation, lead to higher reconstruction errors when the signal is passed through models. This phenomenon makes reconstruction error a reliable indicator for identifying potential pre-ictal activity, serving as an indicator of imminent seizure onset.

In order to investigate these anomalies, three reconstruction-based models were used: the Long Short-Term Memory Autoencoder (LSTM-AE), the Multi-Head Convolutional LSTM Autoencoder (MH-C-LSTM-AE), and the Transformer-Encoder-Encoder (T-EE)-based Anomaly Detection Model. Each model is meticulously crafted to address distinct challenges inherent in ECG signal processing. These challenges include capturing long-term temporal dependencies, modeling complex spatial relationships, and identifying sequential patterns with high fidelity. By independently evaluating these models, the study rigorously evaluates their individual contributions to anomaly detection, providing insights into their effectiveness in advancing seizure prediction.

\subsubsection{Training and Testing Methodology}
In order to develop a stable and personalized seizure prediction framework, a short segment is selected from the initial portion of each patient's ECG recording. This segment represents a typical, non-seizure baseline signal, ensuring that the model is trained exclusively on normal patterns. By focusing on these baseline characteristics, the models develop a comprehensive understanding of the patient's unique ECG dynamics. This enables them to identify subtle deviations from normality during the prediction phase.

Once trained, the models are applied to the entire patient dataset (except the portion selected for training) to identify anomalies indicative of seizure activities. This approach is particularly advantageous for patient-specific analysis, as it inherently accounts for individual variations in ECG morphology and rhythm. By tailoring the models to each patient's data, the framework achieves a high degree of personalization, which is critical for accommodating the diverse physiological characteristics of epileptic patients. Furthermore, the reliance on patient-specific training segments eliminates the need for generalized assumptions, enhancing the performance of anomaly detection \cite{PerezSanchez.2020, PerezSanchez.2024}.

\subsubsection{LSTM Autoencoder}
\begin{figure}[!ht]
	\begin{center}
		\includegraphics[width=\linewidth]{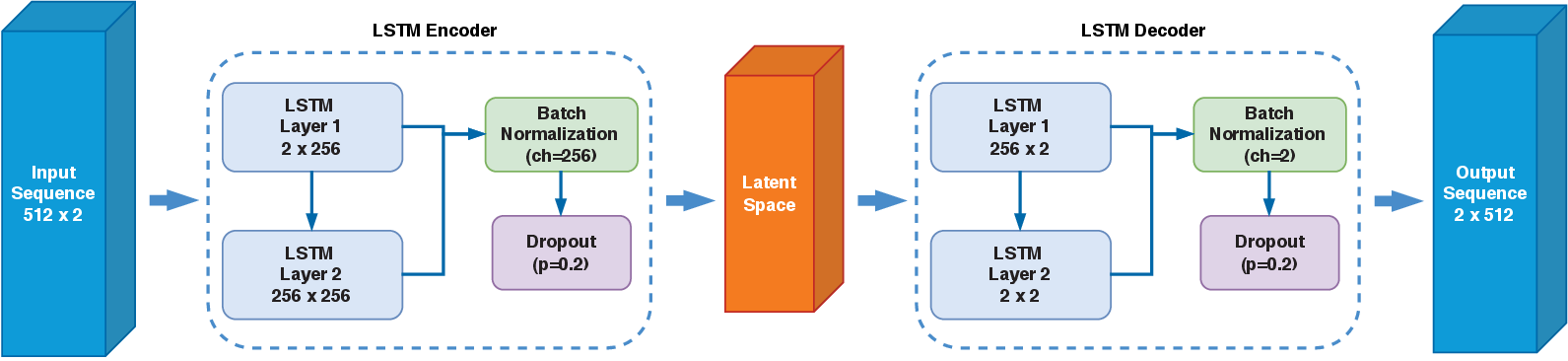}
		\caption{Schematic Diagram of LSTM-AE.}
		\label{fig_lstm}
	\end{center}
\end{figure}

The LSTM-AE is a neural network designed for sequential data, using an encoder-decoder structure. The encoder, composed of stacked LSTM layers, generates a compact latent representation, which captures long-term dependencies within the input data. In order to ensure robust training and to prevent overfitting, optimization layers--Batch Normalization and Dropout--were applied. The decoder, mirroring the encoder, reconstructs the input signal to match the original as closely as possible.
Fig. \ref{fig_lstm} illustrates the LSTM-AE architecture, highlighting its layers and structure.

\subsubsection{Multi-Head Convolutional LSTM Autoencoder}
\begin{figure}[!ht]
	\begin{center}
		\includegraphics[width=\linewidth]{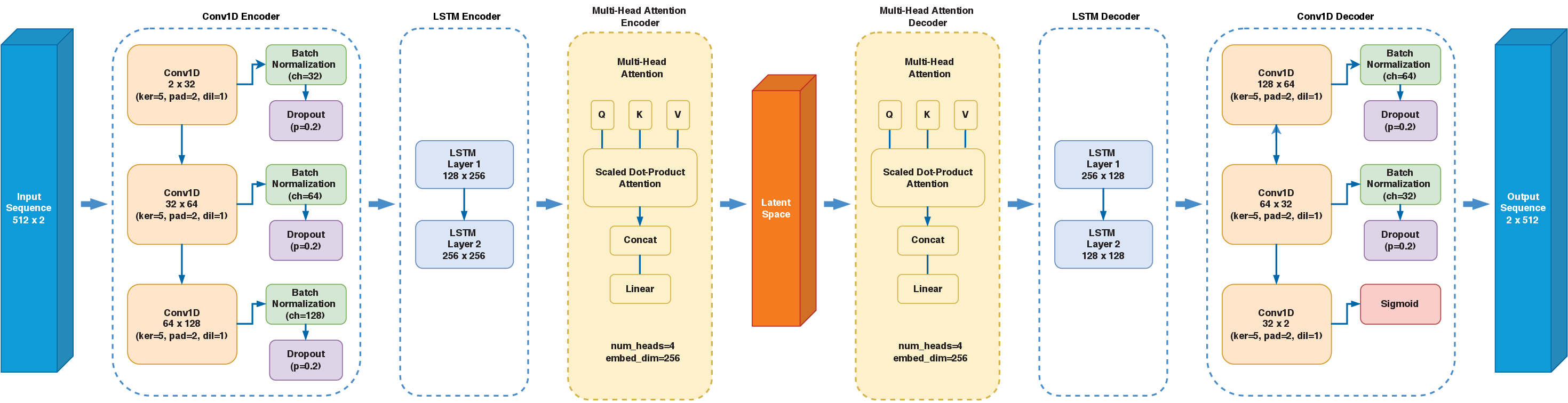}
		\caption{Schematic Diagram of Multi-Head-Conv-LSTM-AE.}
		\label{fig_multi_head}
	\end{center}
\end{figure}

MH-C-LSTM-AE is a hybrid model that integrates convolutional layers, LSTM layers, and multi-head attention mechanisms to effectively capture both spatial and temporal complexity.
In this model, the encoder begins with 1D convolutional layers, which identify localized features such as QRS complexes and waveform patterns. These layers use dilation factors to expand the receptive field, allowing the model to capture both fine-grained and broader patterns. Next, LSTM layers model temporal dependencies, tracking periodic trends and abrupt transitions. To further enhance representation learning, multi-head attention mechanisms are incorporated into the latent space, allowing the model to selectively focus on informative temporal regions and capture long-range dependencies within the signal. This attention mechanism facilitates the identification of subtle abnormalities potentially associated with pre-ictal states, even in the presence of complex background variations.
The decoder reconstructs the input signal through complementary attention, LSTM, and convolutional operations, preserving both temporal and spatial characteristics necessary for accurate reconstruction-based anomaly detection. A sigmoid activation function is applied at the output layer to constrain reconstructed values within the normalized input range. The architecture is motivated by the need to comprehensively capture the multiscale nature of ECG-derived representations, which often contain both localized events (e.g., QRS complexes) and long-term dependencies (e.g., HRV) \cite{Canizo.2019}. The architecture of this model is illustrated in Fig. \ref{fig_multi_head}.

\subsubsection{Transformer-Encoder Encoder}
\begin{figure}[!ht]
	\begin{center}
		\includegraphics[width=\linewidth]{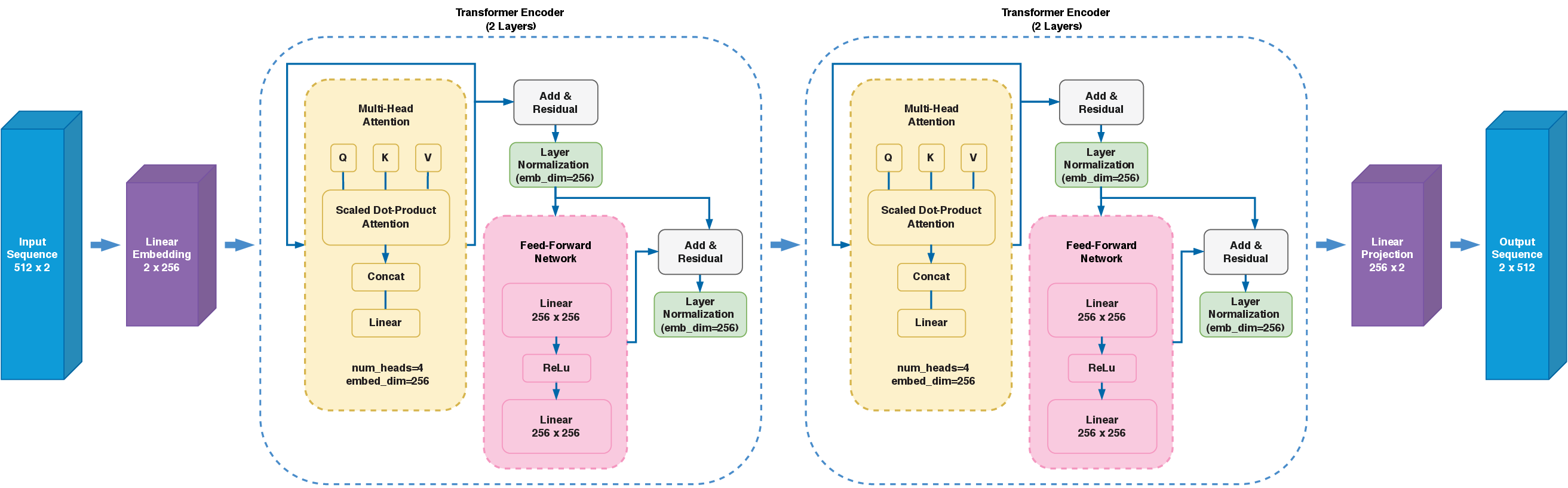}
		\caption{Schematic Diagram of Transformer (Enc-Enc).}
		\label{fig_transformer}
	\end{center}
\end{figure}

Transformers are highly effective for seizure prediction due to their ability to model long-range dependencies in ECG signals. Unlike RNNs and LSTMs, which process data sequentially, transformers operate in parallel, significantly improving efficiency and scalability--critical for analyzing continuous ECG recordings.
The proposed transformer-based model leverages two transformer encoder layers to process ECG signal representations. Each layer employs a multi-head self-attention mechanism, allowing the model to capture key dependencies and relationships within the signal, regardless of their position in the sequence. This capability enables the model to focus on different signal segments at various time steps, offering a flexible and adaptive approach to interpreting complex, non-linear interactions. Following self-attention, a feedforward network refines the learned representations by capturing higher-order feature relationships, further enhancing the model's ability to detect seizure-related patterns. The overall architecture is depicted in Fig. \ref{fig_transformer}.

\section{Post-Processing} \label{sec_postproces}
\subsection{Smoothing}
In the post-processing stage of our seizure prediction framework, smoothing the reconstruction error is a critical step that enhances the robustness and accuracy of predictions. Raw reconstruction error values, derived from comparing the model's output with the original signal, often exhibit significant variability due to noise and transient fluctuations in ECG signal. Without adequate smoothing, these variations could lead to false positives, reducing the predictive accuracy and reliability of the model. In order to resolve this, we apply a moving average technique with a fixed window size, where each point in the smoothed reconstruction error sequence is computed as the mean of its surrounding values. This reduces noise while retaining significant variations associated with seizure events. Specifically, the smoothed reconstruction error $E_{\text{smoothed}}(i)$ at index $i$ is given by:
\begin{equation}
	E_{\text{smoothed}}(i) = \frac{1}{n} \sum_{j=\max(0, i - \frac{w}{2})}^{\min(N, i + \frac{w}{2})} E(j)
\end{equation}
where $E(j)$ represents the raw reconstruction error at point $j$, $w$ is the window size, and $N$ is the total number of points. This technique effectively reduces short-term variability while preserving the overall trend of the reconstruction error curve, which is essential for robust anomaly detection. In order to maintain consistency, we apply smoothing to both the training and testing reconstruction error sequences. Importantly, the smoothed error distribution from the training set serves as the foundation for threshold selection. By leveraging statistical properties of these smoothed errors, we define an adaptive threshold for anomaly detection, ensuring that test samples are evaluated against a stable and noise-filtered baseline.

As a result of this post-processing step, the signal-to-noise ratio of the system is significantly improved, resulting in improved reliability and stability during the prediction phase. It helps to mitigate the risk of false positives caused by transient spikes in error values, ensuring that the model only flags events that exceed a well-defined and smoothed threshold. By integrating this smoothing process, our approach achieves a more balanced performance, critical for real-time seizure prediction applications.

\subsection{Threshold Selection}
The statistical thresholding method employed in this study provides a robust and dynamic mechanism for identifying anomalies by leveraging statistical properties of reconstruction errors. Specifically, this approach calculates the threshold based on the mean and standard deviation of the reconstruction errors derived from the training dataset. This ensures adaptability to patient-specific data distributions and robustness against noise.

The statistical threshold is mathematically defined as follows:
\begin{equation}
	\mu = \frac{1}{N} \sum_{i=1}^{N} E_{\text{Train}}(i), \quad \sigma = \sqrt{\frac{1}{N} \sum_{i=1}^{N} \big(E_{\text{Train}}(i) - \mu\big)^2}
\end{equation}
where $\mu$ represents the mean reconstruction error, and $\sigma$ is the standard deviation of reconstruction errors in the training data. Accordingly, $\tau$ denoted as the threshold is defined as:
\begin{equation}
	\tau = \mu + k\sigma
\end{equation}
where $k$ is a tunable parameter that determines the sensitivity of the anomaly detection system. A higher $k$ reduces false positives by requiring larger deviations to classify an event as anomalous, while a lower $k$ increases sensitivity but may result in more false positives.

In this study, $k$ was conservatively set to 2, striking a balance between minimizing false positives and ensuring the detection of meaningful anomalies.

This method offers significant advantages:

\begin{itemize}
	\item The inclusion of the standard deviation ensures the method is resilient to outliers and irregularities in the training data.
	\item The adjustable $k$-parameter allows flexibility for different application requirements, such as prioritizing high specificity or sensitivity.
\end{itemize}

By employing this statistical thresholding technique, the study ensures that anomalies are detected effectively, providing a foundation for the development of reliable ECG-based seizure prediction systems.

\section{Performance Evaluation} \label{sec_evaluation}
Accurately assessing a seizure prediction model requires a comprehensive evaluation framework that ensures reliability, robustness, and clinical viability. This section outlines the key performance metrics used to quantify predictive accuracy, specificity, and false alarm rates. Additionally, to account for seizure rarity, a class weighting mechanism is applied, preventing model bias toward non-seizure intervals. Another critical aspect is the definition of the pre-ictal interval, which determines the valid window for seizure predictions and ensures a fair assessment across different datasets. By incorporating these evaluation strategies, we establish a well-balanced and interpretable performance analysis.

\subsection{Evaluation Metrics}
In order to fully assess the performance of seizure prediction models, a variety of evaluation metrics should be used. These metrics are based on four fundamental components of a confusion matrix:
\begin{description}
	\item [TP] These are instances where the model correctly predicted seizures.
	\item [FP] These are instances where the model incorrectly predicted a seizure.
	\item [TN] These are the predictions where the model identified normal behavior correctly, avoiding false alarms.
	\item [FN] These are missed seizures, where the model failed to predict the seizure onset.
\end{description}
Seizure prediction models often face a significant class imbalance between seizure intervals (positive class) and normal intervals (negative class). Since seizures are relatively rare compared to normal intervals, this imbalance can distort performance metrics. To address this challenge, a weighting mechanism is employed to adjust the relative importance of each class. Therefore, the weight for the negative class (normal intervals) is set to 1, while the positive class (seizure intervals) is assigned a weight inversely proportional to its frequency. This ensures that the model is penalized more heavily for missing seizures (FN) or falsely predicting seizures (FP), reflecting the critical nature of these errors. Accordingly, the model becomes more sensitive to the minority class, improving its ability to detect seizures without being overly biased by the abundance of normal intervals.

Another vital aspect of seizure prediction evaluation is defining the pre-ictal interval--the period preceding a seizure during which predictions are considered valid. This interval determines the temporal window in which the model's predictions are analyzed. In this study, the pre-ictal interval is dynamically adapted based on the length of the dataset, so for a longer dataset, an hour-long pre-ictal interval is used, while for shorter datasets, a proportionately shorter interval of 30 minutes (half of the standard interval) is applied. Adjusting the pre-ictal interval dynamically ensures that evaluation metrics remain clinically meaningful and consistent across datasets of varying durations. Without this adaptation, models trained on shorter datasets might exhibit artificially high sensitivity, while those trained on longer datasets might fail to detect shorter pre-seizure patterns. This approach offers multiple advantages:

\begin{itemize}
	\item By adjusting the pre-ictal interval to the length of each record, the evaluation ensures that longer or shorter records do not skew performance metrics.
	\item The model is better equipped to handle diverse ECG signal durations, leading to more reliable predictions across varying scenarios.
\end{itemize}

As a consequence, in order to maintain consistency in evaluations, pre-ictal intervals must align with the segmentation window size used in data processing. For instance, if the pre-ictal interval is defined as 3600 samples with a 1-second segmentation window, it must be reduced to 360 samples when a 10-second segmentation window is applied. This adjustment ensures that predictions remain precise and consistent with the scaled data, preventing inaccuracies caused by mismatched intervals and segmentation lengths.
Allowing the above strategies, various evaluation metrics are computed to assess the model's performance:

\begin{itemize}
	\item Accuracy: Represents the proportion of correct predictions (TP and TN) out of all predictions.
	      \begin{equation}
		      \text{Accuracy} = \frac{\text{TP} + \text{TN}}{\text{TP} + \text{FP} + \text{TN} + \text{FN}}
	      \end{equation}

	\item Specificity: Measures the model's ability to correctly identify normal activities (non-seizures).
	      \begin{equation}
		      \text{Specificity} = \frac{\text{TN}}{\text{TN} + \text{FP}}
	      \end{equation}

	\item False Positive Rate (FPR): Indicates the proportion of normal activities incorrectly classified as seizures, helping to assess the model's false alarm rate.

	      \begin{equation}
		      \text{FPR} = \frac{\text{FP}}{\text{FP} + \text{TN}}
	      \end{equation}
\end{itemize}

\section{Experimental Results}\label{sec_results}
In this section, the results obtained through implementation of the proposed framework are presented and analyzed using the evaluation metrics discussed previously. Based on the fundamental steps of the proposed method, the results can be examined from three perspectives: the length of signal segmentation windows, the type of extracted features and representations, and the architecture used for modeling the seizure prediction process. Accordingly, the following analysis of the results is conducted with these considerations in mind.

\subsection{Seizure Prediction Performance}
For a better understanding of the results, it is important to understand how the proposed framework works and what are the preliminary steps to improve prediction performance.

\begin{figure}[!ht]
	\begin{center}
		\includegraphics[width=15cm]{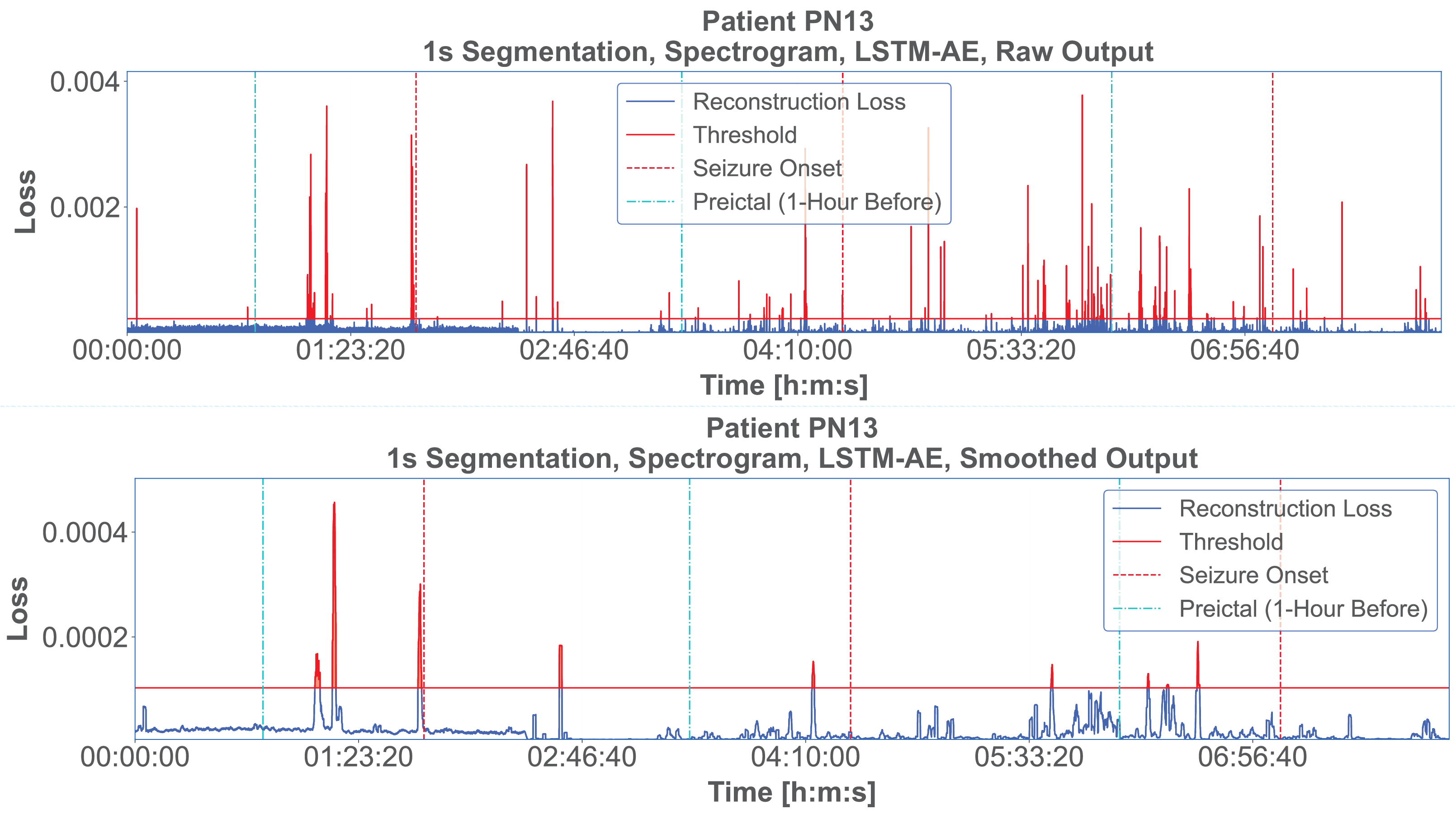}
		\caption{Comparison of Raw and Smoothed Reconstruction Loss (Patient PN13, 1s segmentation, Scalogram, LSTM-AE). In the plots, the blue curve represents the reconstruction error, the red horizontal line indicates the threshold, the red dashed lines mark the seizure onset points, and the turquoise dash-dotted lines, highlight the pre-ictal period.
		}
		\label{fig_raw_vs_smoothed}
	\end{center}
\end{figure}

The proposed framework is designed to predict seizures onset by monitoring and analyzing the reconstruction loss of ECG signals, which serves as an indicator of abnormal patterns in the data. The approach involves extracting meaningful features using time-frequency methods and analyzing them through sequence-to-sequence models. These models are trained to reconstruct ECG signals from normal, non-seizure segments. During this process, the reconstruction loss indicates how well the model can predict the original signal based on its learned patterns. Higher reconstruction errors typically correspond to anomalies, such as those occurring during the pre-ictal phase, which precedes a seizure.
However, as observed in the raw reconstruction error values (Fig. \ref{fig_raw_vs_smoothed}), the presence of noise can lead to false positives, where the model identifies certain points as anomalies, even when they are not indicative of seizure activity. These noise-induced fluctuations complicate the interpretation of the data and decrease the trustworthiness of the predictions.

This is where the post-processing step becomes critical. During this step, a smoothing technique is applied to reduce high-frequency noise in order to allow the model to identify more meaningful anomalies. As shown in the lower part of Fig. \ref{fig_raw_vs_smoothed}, after smoothing, the reconstruction error values are significantly more stable, with fewer false anomalies identified. This improves the model's ability to accurately detect the true pre-ictal phase, reducing the number of incorrect predictions and providing a more reliable prediction method.

\begin{figure}[!ht]
	\vspace{-0.35cm}
	\begin{center}
		\includegraphics[width=15cm]{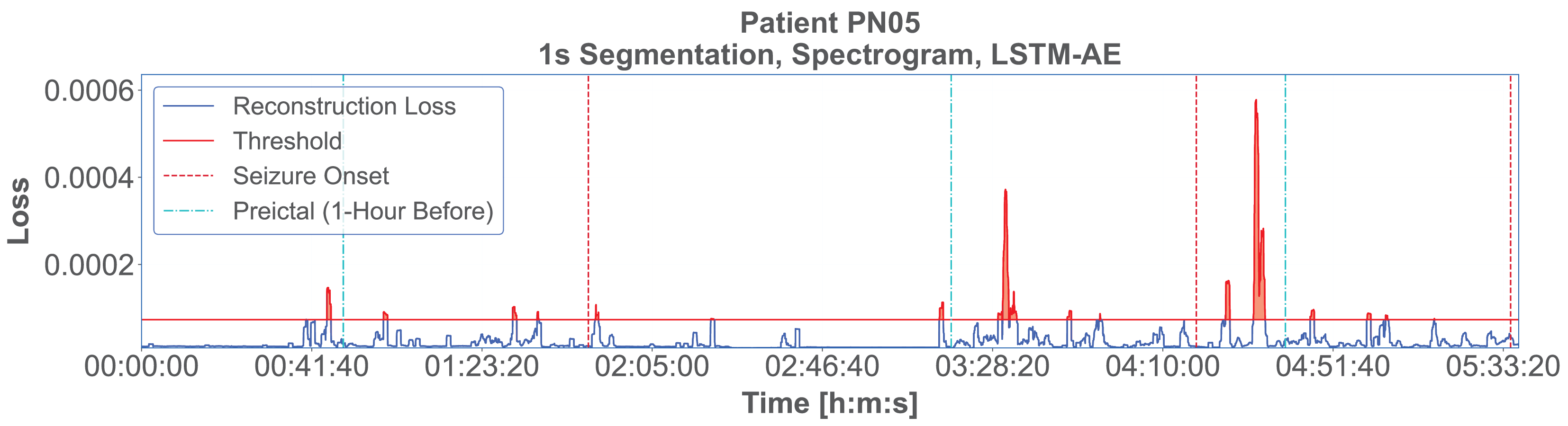}
		\caption{The Reconstruction Loss of Patient PN05 (Patient PN05, 1s segmentation, Scalogram, LSTM-AE). In this case,
			two out of the three seizures were correctly predicted, while the third attack was missed.}
		\label{fig_single_result}
	\end{center}
\end{figure}

In the example shown in Fig. \ref{fig_raw_vs_smoothed}, three out of three seizures were correctly predicted. However, it is important to note that this approach heavily depends on the patient's ECG signal, and in some cases, it may fail to accurately predict impending seizures, as demonstrated in Fig. \ref{fig_single_result}.
In this case, the first two seizure onsets were successfully predicted, while the third attack was missed.

\subsection{Results Based on Segmentation Window Lengths}
The length of segmentation windows is crucial in seizure prediction, directly impacting model performance and reliability. Longer windows capture richer temporal dynamics, enhancing feature stability and accuracy but increasing computational cost and response time. Conversely, shorter windows improve efficiency but may lack sufficient data representation, limiting generalization. This study analyzed segmentation windows of 1, 5, and 10 seconds with varying overlap levels (none, 1, 3, and 5 seconds).

Results showed that a 1-second window without overlap achieved the best results, offering high accuracy and low variability across patients. This setup effectively balances temporal information capture with computational efficiency.
In contrast, the 5-second segmentation window, when used with partial overlaps (e.g., 1 or 3 seconds), showed moderate performance. Although the overlaps enhanced temporal continuity, the results were less consistent than with the 1-second window, as indicated by increased standard deviations. This suggests that while partial overlaps may provide some advantages in terms of continuous data representation, they also introduce more variability, leading to reduced robustness in comparison. Finally, the 10-second segmentation windows, whether used with no overlap or partial overlap, exhibited the lowest performance and highest variability across patients. This performance drop can be attributed to over-aggregation of data, which may obscure the short-term patterns critical for accurate seizure prediction.

\subsection{Results Based on Extracted Feature Types}
The quality and diversity of extracted features are crucial for the success of seizure prediction models. Transforming input ECG signals into meaningful representations enables better analysis of the underlying patterns. In this study, we used three feature extraction methods: DWT, CWT, and STFT, which capture time, frequency, and scale-based features which are useful for predicting seizures.

The performance of each feature extraction method was assessed individually for each patient, and the mean performance across all patients alongside the standard deviation was reported. The standard deviation provides a comprehensive view of the model's robustness across different patients, highlighting variability in performance.
The results are summarized in Tables \ref{tbl_dwt_data}, \ref{tbl_scalogram_data}, and \ref{tbl_spectrogram_data}.
\begin{table}[!ht]
	\centering
	\caption{Summarized results from the DWT representation}
	\label{tbl_dwt_data}
	\renewcommand{\arraystretch}{1.15}
	\scalebox{0.9}{
		\begin{tabular}{c|c|c|c}
			\hline
			Models       & Specificity           & Accuracy               & FPR (/h)              \\ \hline
			LSTM-AE      & 98.16 ± 0.009         & 75.48 ± 0.125          & 0.018 ± 0.009         \\
			MH-C-LSTM-AE & 98.13 ± 0.009         & 75.46 ± 0.125          & 0.019 ± 0.009         \\
			T-EE         & \textbf{98.32 ± 0.01} & \textbf{75.54 ± 0.126} & \textbf{0.017 ± 0.01} \\ \hline
		\end{tabular}
	}
\end{table}
\begin{table}[!ht]
	\centering
	\caption{Summarized results from the Scalogram representation}
	\label{tbl_scalogram_data}
	\renewcommand{\arraystretch}{1.15}
	\scalebox{0.9}{
		\begin{tabular}{c|c|c|c}
			\hline
			Models       & Specificity            & Accuracy               & FPR (/h)              \\ \hline
			LSTM-AE      & 98.8 ± 0.005           & 75.67 ± 0.13           & 0.012 ± 0.005         \\
			MH-C-LSTM-AE & \textbf{98.98 ± 0.006} & \textbf{75.74 ± 0.129} & \textbf{0.01 ± 0.006} \\
			T-EE         & 98.72 ± 0.004          & 75.62 ± 0.13           & 0.013 ± 0.004         \\ \hline
		\end{tabular}
	}
\end{table}
\begin{table}[!ht]
	\centering
	\caption{Summarized results from the Spectrogram representation}
	\label{tbl_spectrogram_data}
	\renewcommand{\arraystretch}{1.15}
	\scalebox{0.9}{
		\begin{tabular}{c|c|c|c}
			\hline
			Models       & Specificity            & Accuracy               & FPR (/h)              \\ \hline
			LSTM-AE      & 98.76 ± 0.008          & 75.95 ± 0.128          & 0.011 ± 0.006         \\
			MH-C-LSTM-AE & \textbf{99.16 ± 0.006} & \textbf{76.05 ± 0.127} & \textbf{0.01 ± 0.005} \\
			T-EE         & 98.76 ± 0.008          & 75.62 ± 0.126          & 0.016 ± 0.009         \\ \hline
		\end{tabular}
	}
\end{table}

\begin{figure}[!ht]
	\begin{center}
		\includegraphics[width=15cm]{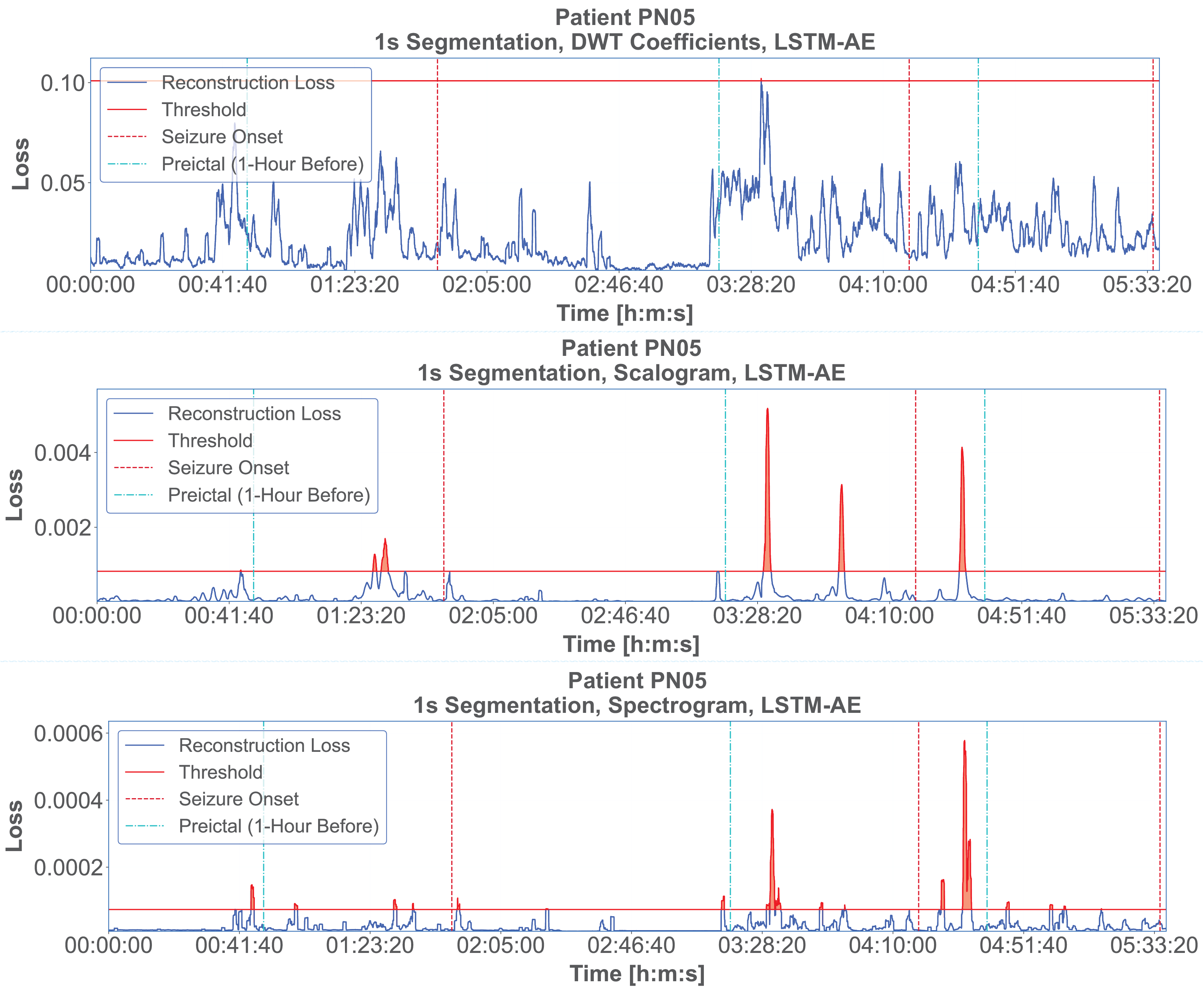}
		\caption{Results of reconstruction loss for seizure prediction (Patient PN05, 1s segmentation, LSTM-AE). The first plot shows the reconstruction loss using combined coefficients from the DWT, the second plot displays results from the scalogram representation derived from the CWT, and the final plot presents the reconstruction loss based on the spectrogram representation from the STFT.}
		\label{fig_result_by_features}
	\end{center}
\end{figure}

Analysis shows that the STFT-derived spectrogram slightly outperforms other methods, suggesting that frequency-domain information offers more discriminative features for seizure prediction. However, as illustrated in Fig. \ref{fig_result_by_features}, scalogram and spectrogram, displayed similar trends, indicating that pre-seizure anomalies occurred at consistent intervals. These points could potentially aid in early seizure detection, though further clinical validation is needed.

In conclusion, while all feature extraction methods show potential, the STFT-based spectrogram offers the most promising results for seizure prediction, guiding future research in selecting the most effective techniques.

\subsection{Results Based on Designed Models}
In the final step, architectures play a crucial role in transforming the extracted features into actionable predictions. In this study, a wide range of neural network architectures were employed, including feedforward networks, convolutional networks, recurrent networks, autoencoders, and transformer models. The results of these models are presented cumulatively in Table \ref{tbl_final_results}.

\begin{table}[!ht]
	\centering
	\caption{Final Results}
	\label{tbl_final_results}
	\renewcommand{\arraystretch}{1.15}
	\scalebox{0.9}{
		\begin{tabular}{c|c|c|c}
			\hline
			Models       & Specificity            & Accuracy               & FPR (/h)              \\ \hline
			LSTM-AE      & 98.76 ± 0.008          & 75.95 ± 0.128          & 0.011 ± 0.006         \\
			MH-C-LSTM-AE & \textbf{99.16 ± 0.006} & \textbf{76.05 ± 0.127} & \textbf{0.01 ± 0.005} \\
			T-EE         & 98.76 ± 0.008          & 75.62 ± 0.126          & 0.016 ± 0.009         \\
			\hline
		\end{tabular}
	}
\end{table}

Based on the data shown in Table \ref{tbl_final_results}, it is evident that among the proposed architectures, the MH-C-LSTM-AE model achieved the highest performance, successfully predicting 45 out of 47 seizures with specificity of 99.16\%, accuracy of 76.05\%, and FPR of 0.01/h. Following closely, the T-EE model also demonstrated high performance, predicting 44 seizures, and finally, the LSTM-AE model predicted 43 seizures.
Additionally, it can be observed that all three models were able to predict seizures approximately 40 minutes in advance, indicating their potential for early seizure prediction.

\section{Comparison with Previous Studies} \label{sec_previous_works}
Seizure prediction using ECG signals has been extensively investigated in biomedical signal processing, with approaches evolving from statistical analysis of HRV to more recent learning-based and anomaly detection frameworks. To provide a methodical comparison with current approaches., Table \ref{tbl_tradeoffs} summarizes representative ECG-based studies, highlighting differences in modeling strategies, prediction horizons, and reported performance.
Early works, such as Yamakawa et al. \cite{Yamakawa.2020}, demonstrated the feasibility of real-time seizure prediction using wearable ECG systems, while clustering-based approaches \cite{Gagliano.2020, Leal.2021} provided initial evidence that pre-ictal states can be identified through changes in HRV dynamics. However, these methods rely primarily on handcrafted statistical features, which limit their ability to capture the non-linear and non-stationary nature of cardiac activity preceding seizures.

More recent studies have shifted toward anomaly detection and deep learning-based frameworks. In particular, Ode et al. \cite{Ode.2022} proposed a reconstruction-based self-attentive autoencoder operating on RR interval signals, demonstrating that reconstruction error can effectively capture pre-ictal deviations. Building on this paradigm, subsequent studies have explored both classical and learning-based anomaly detection strategies, as well as improved feature representations for enhancing generalization. Despite these advances, most existing approaches remain constrained by their reliance on HRV/RRI-based representations or fixed feature extraction pipelines, which may not fully capture the multi-scale temporal and spectral characteristics of ECG signals.

Motivated by these limitations, the proposed framework integrates time-frequency representations with reconstruction-based deep architectures to enable modeling of ECG dynamics across both temporal and spectral domains. This allows the capture of both transient variations and longer-term autonomic patterns associated with pre-seizure states. As summarized in Table \ref{tbl_tradeoffs}, existing methods often exhibit limitations in predictive performance and robustness, particularly in terms of FPR, which remains a critical challenge for clinical applicability. By leveraging these representations within a deep anomaly detection framework, the proposed method addresses these shortcomings, resulting in improved robustness--most notably through reduced FPR--and thereby enhancing the reliability and practical suitability of ECG-based seizure prediction systems.


\begin{table}[ht]
	\caption{Recent ECG-based seizure prediction studies using anomaly detection approach.}
	\label{tbl_tradeoffs}
	\begin{adjustbox}{width=\textwidth}
		\begin{tabular}{c|c|ccc|c|>{\arraybackslash}m{0.5\textwidth}}			\hline
			\multirow{2}{*}{\textbf{Study}}                & \multirow{2}{*}{\textbf{Method}} & \multicolumn{3}{c|}{\textbf{Performance}}            & \multirow{2}{*}{\textbf{\makecell{Prediction                                                                                                                                                                                                         \\ Horizon}}}   & \multirow{2}{*}{\textbf{Key Contribution}}                                                                                                                                                                                                \\ \cline{3-5}
			                                               &                                  & \multicolumn{1}{c|}{\textbf{\makecell{Specificity}}} & \multicolumn{1}{c|}{\textbf{\makecell{Sensitivity}}} & \textbf{\makecell{FPR}}                                                        &                   &                                                                                          \\ \hline
			Yamakawa et al. \cite{Yamakawa.2020}           & MSPC                             & \multicolumn{1}{c|}{-}                               & \multicolumn{1}{c|}{85.7\%}                          & 0.62                                                                           & About 5 min       & Demonstrated feasibility of real-time seizure prediction using wearable ECG systems.     \\ \hline
			Gagliano et al. \cite{Gagliano.2020}           & K-Means clustering               & \multicolumn{1}{c|}{-}                               & \multicolumn{1}{c|}{-}                               & -                                                                              & \makecell{Between                                                                                            \\ 3.5 and 6.5 min} & Revealed early pre-ictal clustering patterns associated with HRV dynamics.                      \\ \hline
			Leal et al. \cite{Leal.2021}                   & Clustering algorithms            & \multicolumn{1}{c|}{-}                               & \multicolumn{1}{c|}{-}                               & -                                                                              & Up to 40 min      & Showed heterogeneous patient-specific emergence of pre-ictal patterns.                   \\ \hline
			Ode et al. \cite{Ode.2022}                     & Self-Attentive AE                & \multicolumn{1}{c|}{-}                               & \multicolumn{1}{c|}{74\%}                            & 0.85                                                                           & -                 & Introduced reconstruction-based anomaly detection using attention-enhanced autoencoders. \\ \hline
			Karasmanoglou et al. \cite{Karasmanoglou.2023} & \makecell{One-Class SVM                                                                                                                                                                                                                                                                                                                        \\ MCD\\ LOF}     & \multicolumn{1}{c|}{\makecell{93.1\%\\ 87.8\%\\ 96.6\%}}             & \multicolumn{1}{c|}{\makecell{\textbf{95.6\%}\\ 91.1\%\\ 92.4\%}} & -                                                                              & \makecell{Between \\ 6 and 30 min}    & Provided comparative evaluation of classical anomaly detection methods for pre-ictal detection. \\ \hline
			Behbahani et al. \cite{Behbahani.2024}         & Thresholding                     & \multicolumn{1}{c|}{-}                               & \multicolumn{1}{c|}{\makecell{80.42\%                                                                                                                                                                                                                \\ 75.19\%}} & 0.15                                                                           & -                                                             & Modeled temporal autonomic dynamics using multi-lag Poincaré representations.            \\ \hline
			Jiang et al. \cite{Jiang.2024}                 & MSPC                             & \multicolumn{3}{c|}{Accuracy of 88.2\%}              & -                                                    & Improved generalization by incorporating broader feature sets across patients.                                                                                                                \\ \hline
			Proposed Method                                & \makecell{LSTM-AE                                                                                                                                                                                                                                                                                                                              \\ MH-C-LSTM-AE\\ T-EE} & \multicolumn{1}{c|}{\makecell{98.76\%\\ \textbf{99.16}\%\\ 98.76\%}} & \multicolumn{1}{c|}{\makecell{-\\ -\\ -}}         & \makecell{0.011\\ \textbf{0.010}\\ 0.016}         & \makecell{43\\ \textbf{45}\\ 44} & Hybrid multi-architecture anomaly detection using time-frequency ECG representations.    \\ \hline
		\end{tabular}
	\end{adjustbox}
\end{table}

\section{Discussion}\label{sec_discussion}
Seizure prediction remains a critical challenge in neurology, with significant implications for patient safety and quality of life. The findings of this study provide strong evidence for the effectiveness of an ECG-based anomaly detection framework for seizure prediction, with high accuracy and temporal sensitivity. The framework, grounded in time-frequency domain feature extraction and advanced sequence modeling coupled with reconstruction error analysis, enables the proposed framework to not only detect seizures accurately but also provide a dynamic, patient-specific solution using adaptive statistical thresholds. This is critical for real-world clinical applications, where individual variability in ECG signals often challenges the use of universal thresholds.

One of the critical elements contributing to the model's success was the smoothing of reconstruction errors using moving average filtering. This step was pivotal in enhancing the reliability of the model by effectively reducing transient fluctuations often inherent in ECG signals. By reducing these fluctuations, the model is able to maintain accuracy while decreasing false alarms, which is essential for clinical systems. Furthermore, the use of an adaptive statistical thresholding method, derived from the training set's distribution parameters, enables the creation of patient-specific decision boundaries. This method accounts for inter-patient variability in physiological signals, ensuring that the model can effectively adapt to the individual characteristics of each patient.
Next, the choice of segmentation window length also played a crucial role in model performance. In particular, a 1-second window without overlap provided superior results, striking a balance between temporal granularity and computational efficiency. This setup allowed the model to capture fine-grained temporal features, which are essential for detecting the pre-ictal state. Longer windows, conversely, resulted in decreased performance, due to over-smoothing and the loss of temporal resolution, which is critical for identifying early pathological changes.
The feature extraction methodology also influenced the system's accuracy. Among the methods evaluated, STFT demonstrated the highest discriminative capability, highlighting the importance of frequency-domain features in identifying pre-seizure ECG patterns. However, the relatively slight performance margin between STFT, CWT, and DWT suggests that a multi-representation ensemble could be a promising avenue for future research. Combining these techniques may enhance the model's ability to detect diverse pre-seizure signatures, potentially improving prediction accuracy.

Finally, among the investigated models, the MH-C-LSTM-AE model achieved the best performance. It successfully identified 45 out of 47 seizures, exhibiting high specificity and FPR of only 0.01/h. The architecture's ability to integrate temporal dependencies (via LSTM layers), spatial hierarchies (via convolutional layers), and contextual attention (via multi-head attention) was crucial for capturing the complex ECG dynamics associated with seizure onset. Notably, the approach predicted seizure onset with an average prediction time of approximately 40 minutes, suggesting its potential for real-time, anticipatory intervention in clinical settings.

Despite these successes, the proposed framework exhibits comparatively lower sensitivity, indicating that a small number of seizures were not detected. This limitation is reflected in the model's conservative detection strategy, which prioritizes minimizing false positives through adaptive thresholding and reconstruction error scoring. While this design choice is advantageous for clinical applicability, where excessive false alarms can reduce usability and patient trust, it can lead to missed detections in cases where pre-ictal patterns are subtle or highly variable across individuals. This trade-off between sensitivity and FPR is a well-recognized challenge in seizure prediction systems. Furthermore, the inter-individual variability in ECG dynamics complicates this issue, limiting the model's ability to generalize across diverse patient profiles.

Furthermore, the proposed methodology has potential for broader clinical applications beyond seizure prediction. Given its reliance on ECG signals and reconstruction-based anomaly detection, the framework may be adapted for the detection and monitoring of other physiological conditions exhibiting temporal abnormalities in cardiac activity. In particular, applications such as arrhythmia classification \cite{Shah.2025} and sleep apnea detection \cite{Chen.2025} represent promising directions, as these conditions share similar time-dependent patterns in ECG signals and related biosignals.

\section{Conclusion}\label{sec_conclusion}
In this study, we developed a comprehensive ECG-based framework for seizure prediction, integrating time-frequency representations with reconstruction-based deep learning architectures. The proposed framework demonstrated high specificity and low FPR, supporting its suitability for real-world clinical deployment where minimizing false alarms is critical. However, this performance is compromised by a trade-off in sensitivity, reflecting the inherent challenge of detecting highly variable pre-ictal patterns across individuals. Moreover, the generalizability of the model is constrained by limited patient-specific data and the difficulty of precisely characterizing seizure phases.
Moving forward, future research should focus on improving sensitivity through approaches such as patient-specific calibration, hybrid decision mechanisms, and transfer learning strategies that better capture individualized pre-ictal signatures. In addition, prospective real-time evaluation is needed to assess whether the trade-off between sensitivity and false alarms remains clinically acceptable in real-world settings.


\newpage
\bibliography{refrences}
\end{document}